\documentclass{article}

\usepackage{amsmath}
\usepackage{amssymb}
\usepackage{graphicx}
\usepackage{multirow}
\usepackage{rotating}
\usepackage{geometry}
\usepackage{color}
\usepackage{authblk}
\usepackage{hyperref}
\usepackage{ulem}

\title{A new type of window functions constructed with exponential function}
\author{Haichao Xu and Xingpao Suo \\ email: \href{Haichao_XU@zju.edu.cn}{Haichao\_XU@zju.edu.cn}}
\affil{Institute for Astronomy, School of Physics, Zhejiang University, 866 Yuhangtang Rd, Hangzhou, 310058, China}

\begin{document}

\maketitle

\begin{abstract}
  The Discrete Fourier Transform (DFT) is widely utilized for signal analysis but is plagued by spectral leakage, leading to inaccuracies in signal approximation. Window functions play a crucial role in mitigating spectral leakage by providing weighting mechanisms for discrete signals. In this paper, we introduce a novel window type based on exponential function, allowing for adjustable parameters and diverse variations. We present the formulation, properties, and motivation behind the design of the new window functions. Additionally, we analyze their behavior and evaluate their performance by comparing them with mainstream window functions using six parameters. Our findings demonstrate that these new window functions exhibit outstanding flexibility and versatility in signal analysis.
\end{abstract}


\section{Introduction}

The Discrete Fourier Transform (DFT) and its fast implementation, the Fast Fourier Transform (FFT)\cite{cooley1965algorithm}, have become indispensable tools for signal analysis in various fields. However, the application of DFT or FFT to discrete signal points suffers from spectral leakage, which hampers the accurate approximation of the true signal. Spectral leakage arises due to the discontinuity and periodic reconstruction of the signal in the time domain. Consequently, signals with inaccurate periods inevitably experience leakage due to the Gibbs phenomenon in mathematics\cite{bocher1906introduction,hewitt1979gibbs}. This issue is particularly challenging in low signal-to-noise ratio (SNR) scenarios, such as gravitational wave signals\cite{klimenkoWavescanMultiresolutionRegression2022,theligoscientificcollaborationGuideLIGOVirgoDetector2020}.

Window functions play a crucial role in signal processing as they help mitigate the problem of spectral leakage\cite{harrisUseWindowsHarmonic1978}. These functions provide a mechanism for weighting discrete signals, effectively reducing the signal intensity near the boundaries of the sampling window and alleviating the issue of periodic discontinuity in the signal. It's important to note that window functions have broader applications beyond spectral analysis alone, such as data smoothing, digital filter design and statistical analysis. Consequently, the search for well-suited window functions with excellent performance remains an important focus in signal analysis.

In this paper, we propose a novel window type based on exponential function. By adjusting the different parameters, we can obtain different variations of the window function. Additionally, we can optimize existing window functions to improve their smoothness and continuity using our reconstruction scheme. In section \ref{sec:theory}, we provide a comprehensive introduction to this type of new window functions, discussing their formulation, properties, and the underlying design motivation. In section \ref{sec:analysis}, we analyze the behavior of the window functions in the frequency domain using FFT by sampling to $t=100$. We evaluate their performance with six parameters and compare them with mainstream window functions.

\section{Motivation and Design Guideline}
\label{sec:theory}

Window functions can be broadly categorized into two types: piecewise and continuous. Piecewise functions define the window within distinct intervals, potentially leading to discontinuities at the endpoints of these intervals. On the other hand, continuous functions offer the advantage of smoothness and avoid abrupt changes. Therefore, when aiming to construct a high-performing window function, we begin with a continuous function as the foundation.

Among continuous window functions, different types can be categorized based on their mathematical form. For example, the Hann window\cite{harrisUseWindowsHarmonic1978} serves as a good example of cosine-type functions. These functions are well-known for their smoothness and continuity, making them effective in reducing spectral leakage during signal processing\cite{sabin2008discrete,proakis2007digital}. Polynomial-type windows, such as the Rectangular and Welch windows\cite{welch1967use}, are based on polynomial functions. While these windows have simple forms, they may not effectively reduce spectral leakage compared to other types\cite{sabin2008discrete}. As an example of bell-shaped distribution function, Gaussian windows\cite{harrisUseWindowsHarmonic1978,achieser2013theory} often require truncation or the application of another window function with an end of 0 to avoid artifacts.

We can enhance window functions by incorporating existing windows with different weight, such as the Blackman window\cite{blackman1958measurement}. By adjusting the weights assigned to different terms in the summative window function, we can modify its performance. However, the fundamental properties of the window function remain dependent on the constituent terms. 

In our pursuit of practicality and efficiency, we want to construct a novel type of window functions that resemble cosine-type functions. These functions should possess a simple form and exhibit excellent continuity and we also want them to serve as a foundation for constructing combined window functions.

For simplicity, we define the window function $W(t)$ in this paper on the interval $(0,1)$ and set it to zero outside this interval. By replacing $t$ with $n_i/N$, we can seamlessly transition to the case of discrete points. From a practical standpoint, it is preferable for $W(t)$ to rapidly converge to 0 at the endpoints of the interval and demonstrate good continuity.

In order to ensure excellent continuity of the window function, we desire its derivative of any order to converge rapidly to 0 at the endpoints 0 and 1. To achieve this, we can construct a type of function using the exponential function as follows:
\begin{equation}
  W(t) = 
  \begin{cases}
    e^{-A(t)}, & 0 < t < 1, \\
    0, & \text{otherwise},
    \end{cases}
\end{equation}
where $A(t)$ is any function that exhibits sufficient continuity and tends to $\infty$ rapidly from a direction where $\mathrm{Re}A(t) > 0$ as $t$ approaches 0 and 1 within the interval $(0,1)$.

We can generate such a function $A(t)$ from another function $B(t)$, i.e.,
\begin{equation}
  A(t) = \frac{1}{B(t)}.
\end{equation}
In order to ensure the desired properties, we require that for small values of $\epsilon$, the following conditions hold:
\begin{equation}
  B(\epsilon),B(1-\epsilon)>0
\end{equation}

There are various functions that satisfy these conditions and can be used as $B(t)$, such as $\ln{\left[\frac{5}{4} - (t-\frac{1}{2})^2\right]}$, $\sin(\pi t)$, and others. It is also possible for $B(t)$ to be a summation window function, which provides great convenience in constructing $B(t)$. It can be proven that for window functions of this form, both $W(t)$ and its derivative of any order are continuous on $\mathbb  R$ and converge rapidly to zero at the endpoints 0 and 1, provided that $B(t)$ or the signals $x(t)$ exhibit moderate variation.

Indeed, any function that satisfies the given conditions can be optimized in this manner to obtain a smoother and more continuous window function. By considering any $B(t)$ that meets the conditions, the reconstructed window function can be expressed as:
\begin{equation}
  W(t) =
  \begin{cases}
  \exp{\left[\frac{1}{B_{\text{max}}}-\frac{1}{B(t)}\right]}, & 0 < t < 1, \\
  0, & \text{otherwise}\ ,
  \end{cases}
\end{equation}
where $B_{\textit{max}}$ is set to the maximum of $B(t)$ on the interval $(0, 1)$ for normalization. 
Among the window functions that meet our requirements, its distribution in the Fourier domain is given by:
\begin{equation}
  \widehat{W}(\omega) = \mathcal{F}(W)=\int_{-\infty}^{\infty} e^{i\omega t} W(t)dt = \int_0^1 e^{i\omega t} W(t)dt.
  \label{eq:FT_integrate}
\end{equation}
Note that a window function with exponent form  has been constructed in \cite{avciNewWindowBased2008a,1163349} as follows  
\begin{equation}
  W(t)=\frac{   \exp \left( \alpha \sqrt{1-4(t-0.5)^2} \right) }{\exp(\alpha)}
\end{equation}
which does not converge to 0 at $t=0,1$.

\begin{figure}[htb]
  \centering
  \includegraphics[height=0.30\textwidth]{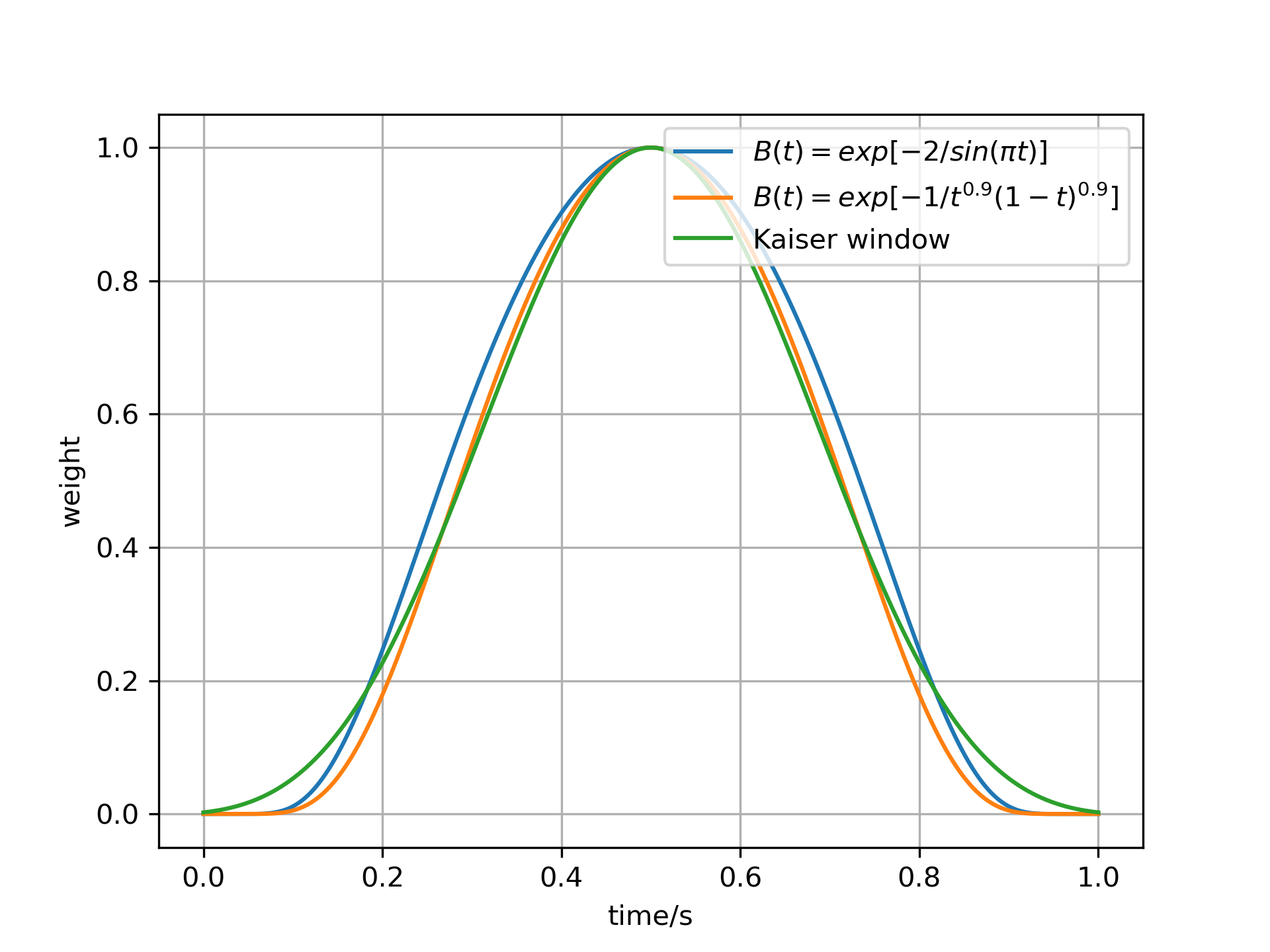} 
  \caption{Approximating the shape of the Kaiser window using specific $B(t)=t^{0.9}(1-t)^{0.9}$ and $\sin(\pi t)/2$.}
  \label{fig:approximation}
\end{figure}

The case where $B(t)$ is a polynomial of the form is of particular interest to us (without normalization)i.e.
\begin{equation}
  B(t) = t^m(1-t)^n
\end{equation}
In this scenario, both $m$ and $n$ are greater than 0. Functions constructed using polynomials offer the advantage of approximating more complex functions by adjusting the exponents $m$ and $n$. For instance, the Kaiser window\cite{1163349} can be approximated using a specific polynomial with $m=n=0.9$ (as shown in Fig. \ref{fig:approximation}).

\begin{figure}[htb]
  \centering
  \includegraphics[height=0.30\textwidth]{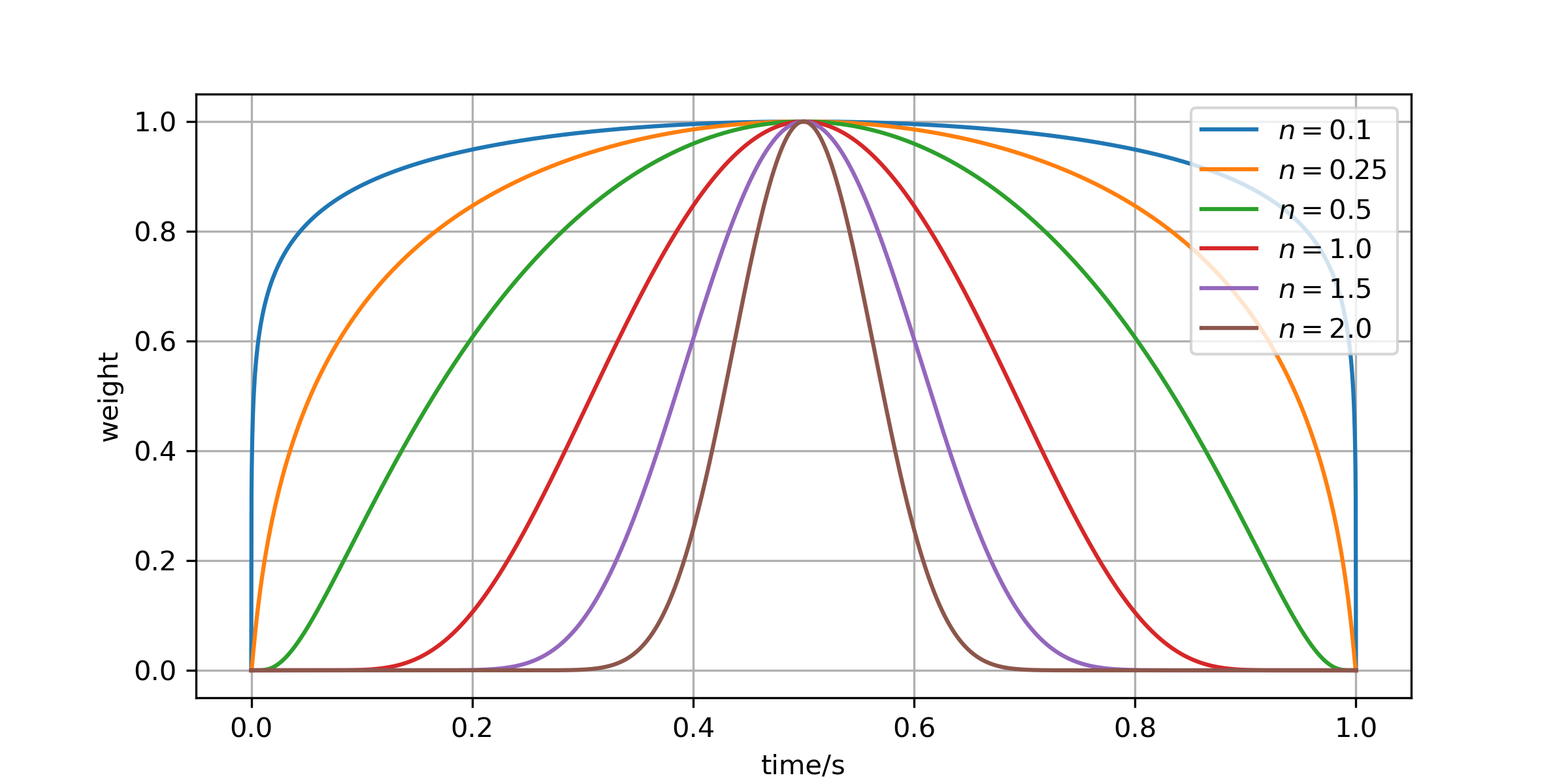} 
  \caption{Symmetric windows constructed using the polynomial $B(t)=t^n(1-t)^n$ for different values of $n$.}
  \label{fig:symmetric_polynomial}
\end{figure}

In certain situations, biased window functions may be required for different purposes. For the constructed window function, when $m=n$, the resulting window function is symmetric over the interval $(0,1)$, whereas when $m\neq n$, we obtain asymmetric window functions.  In the case of symmetry, the window function can be normalized using its value at the midpoint of the interval, i.e
\begin{equation}
  W(t) = 
  \begin{cases}
    \frac{\exp{(4^n)}}{\exp{\left[1/t^n(1-t)^n\right]}}, & 0 < t < 1, \\
    0, & \text{otherwise}.
    \end{cases}
\end{equation}
In Fig. \ref{fig:symmetric_polynomial}, we illustrate some examples of such windows for different values of $n$.

\begin{figure}[htb]
  \centering
  \includegraphics[height=0.30\textwidth]{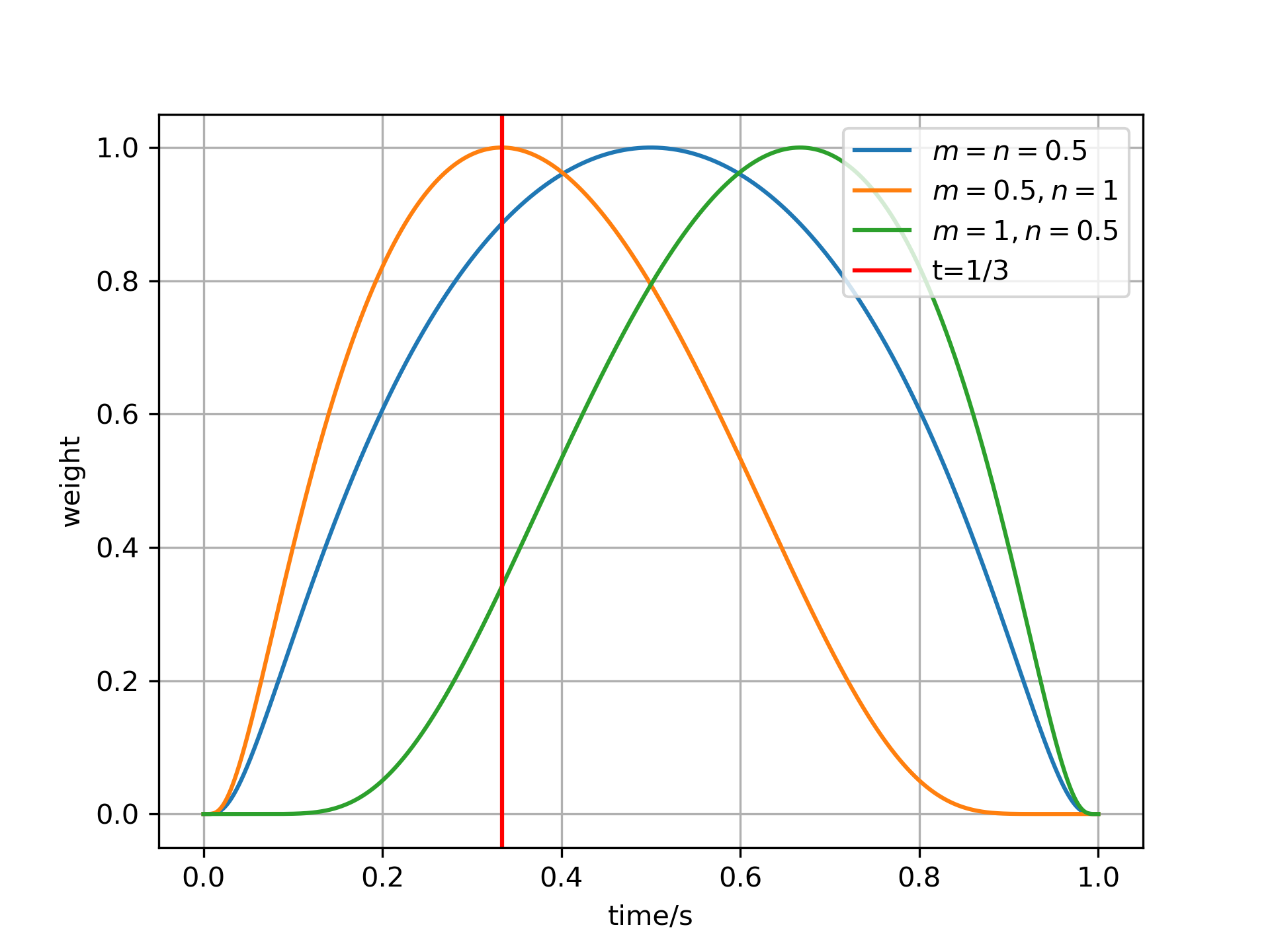}
  \caption{Comparison of symmetric and asymmetric windows constructed with polynomial $B(t)=t^m(1-t)^n$.} 
  \label{fig:asymmetric}
\end{figure}

To determine the maximum value of the function $\exp{\left[-1/t^m(1-t)^n\right]}$ on the interval $(0,1)$ for the asymmetric case, we can calculate its derivative on the interval, which is
\begin{equation}
\frac{m/t-n/(1-t)}{t^m(1-t)^n}\exp{\left[-\frac{1}{t^m(1-t)^n}\right]}.
\end{equation}
The derivative is equal to zero only when $t=m/(m+n)$, indicating the critical point within the interval. Therefore, for the asymmetric case, the normalized window function is given by:
\begin{equation}
W(t) =
\begin{cases}
\frac{\exp{\left[  \left(\frac{m+n}{m}\right)^m  \left(\frac{m+n}{n}\right)^n   \right]}}{\exp{\left[1/x^n(1-x)^n\right]}}, & 0<t<1, \\
0, & \text{otherwise}.
\end{cases}
\end{equation}
In Fig. \ref{fig:asymmetric}, we present some examples of asymmetric windows for different values of $m$ and $n$.

However, since in most cases we require a symmetric window function, we will mainly focus on the case where $m=n$ in the following analysis when $B(t)$ refers to polynomial. Other window functions with simple form and their variants will alse be analyzed as a representative of non-polynomial functions.

\begin{figure}[htb]
  \centering
  \includegraphics[height=0.30\textwidth]{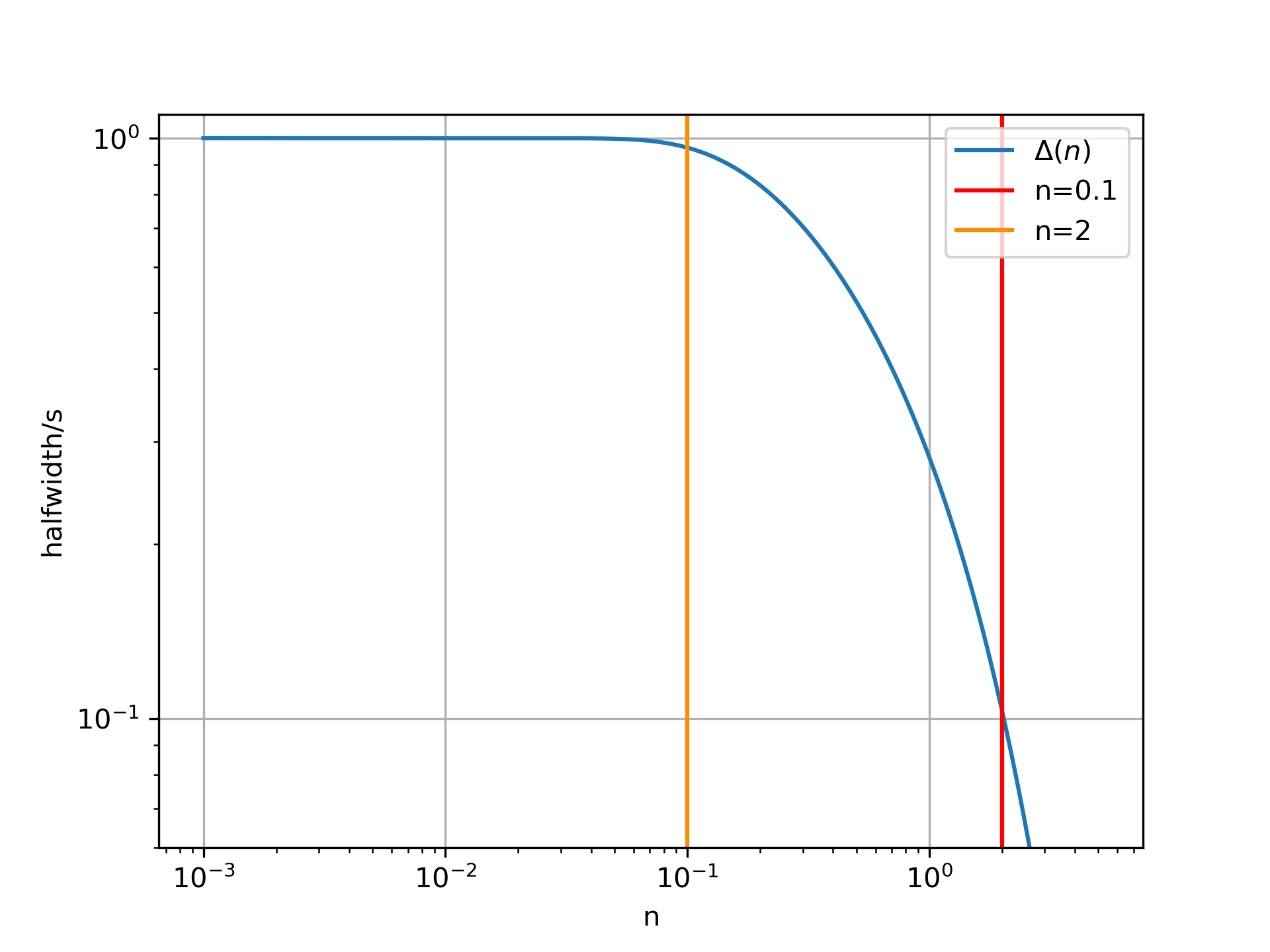} 
  \caption{Change in half width with different power values of $n$.}
  \label{fig:halfwidth}
\end{figure}

Now we analyze how the shape of the symmetric window function changes with different values of $n$. We can define the half width of the window function as the interval length where the energy of the signal decays to less than half of its original value, i.e. $W(t)<\sqrt{2}/2$. By substituting our window function definition into the equation, we can solve for its half width $\Delta$ as a function of $n$
\begin{equation}
\Delta = \sqrt{1 - 4\left( \frac{1}{4^n + \ln{\sqrt{2}}}\right) ^{1/n}}
\end{equation}
From Fig. \ref{fig:halfwidth}, we can observe that when $n > 2$, the half width of the window function becomes very small, approaching a pulse shape. In this case, the window function essentially provides no windowing effect. On the other hand, when $n < 0.1$, the shape of the window function resembles a rectangular window, causing less modification to the signals. However, it is important to note that the window function we defined maintains sufficient continuity at the interval endpoints. Therefore, as $n$ approaches 0, it can achieve functionality similar to a rectangular window while still providing some suppression of spectral leakage.

\section{Performance Analysis}
\label{sec:analysis}

\begin{figure}[htb]
  \centering
  \includegraphics[width=0.6\linewidth ]{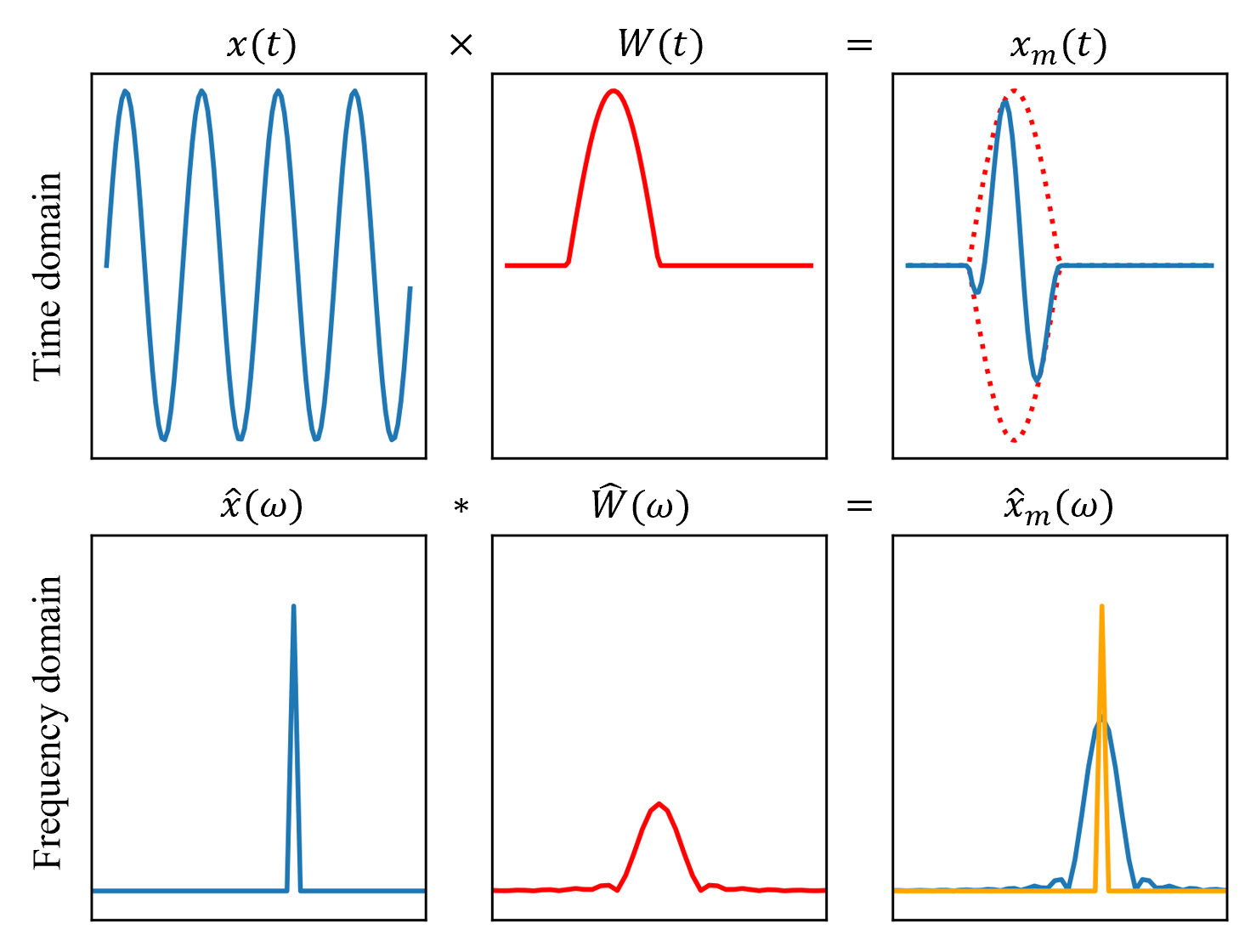} 
  \caption{Modulation effect of the window function on the signal in the time domain and frequency domain.}
  \label{fig:modulation}
\end{figure}

When we apply a window function $W(t)$ defined on the interval $(a,b)$ to a sequence of signals $x(t)$, we obtain a sequence of modulated signals $x(t)\cdot W(t)$. Outside the interval $(a,b)$, the signal strength decays to zero. In the frequency domain, this corresponds to the convolution of the Fourier transforms of the signals\cite{harrisUseWindowsHarmonic1978}, i.e., $\widehat{x}(\omega)*\widehat{W}(\omega)$, as shown in Fig. \ref{fig:modulation}.

Later, when talking about normalized windows, we apply the logarithm of the normalized amplitude spectrum in decibels $dB$, i.e.
\begin{equation}
  \widehat{W}/dB=20\log_{10}\left( \frac{\widehat{W}(\omega)}{\widehat{W}(0)}\right) .
\end{equation}
For a normalized window function, we can analyze its performance using the following well-defined parameters:
\begin{itemize}

  \item Main lobe width $2\Omega_0$. Unlike most other papers, we define $2\Omega_0$ as twice the frequency corresponding to where amplitude $\widehat{W}(\omega)$ decays to 0 for the first time, i.e. 
  \begin{equation}
    \widehat{W}(\Omega_0)=\widehat{W}(-\Omega_0)=0.
  \end{equation}
  This parameter characterizes the extent to which the signal is broadened in the frequency domain and represents the width of the main lobe of the spectrum.

  \item Energy leakage intensity $1-I_0$. Here $I_0$ represents the ratio of the energy gathered at the main lobe to the total energy, i.e
  
  \begin{equation}
    I_0=\frac{\int_{-\Omega_0}^{\Omega_0} \left|\widehat{W}(\omega)\right|^2 d\omega}{\int_{-\infty}^{\infty} \left|\widehat{W}(\omega)\right|^2 d\omega}.
  \end{equation}

  According to Parseval's theorem, the total energy corresponding to the denominator can be simplified using the integral in the time domain, i.e.

  \begin{equation}
    I_0=\frac{\int_{-\Omega_0}^{\Omega_0} \left|\widehat{W}(\omega)\right|^2 d\omega}{\int_0^1 \left|W(t)\right|^2 dt}.
  \end{equation}

  In most cases, $I_0$ is typically similar in meaning to $2\Omega_0$ defined above. However, there are some cases where two different window functions may have very close main lobe widths but different main lobe decay speeds. In such cases, the introduction of this parameter becomes valuable. Additionally, $I_0$ provides a more intuitive representation of the amount of spectral leakage. Since $I_0$ is very close to 1 for good window functions, the calculation of $1-I_0$ offers better intuition.

  \item First sidelobe height $\widehat{W}_1\text{~in~}dB$. When a window function is applied, the leakage is mainly concentrated in the first sidelobe. This parameter indicates the height of the first sidelobe relative to the main lobe and quantifies the maximum height of spurious peaks caused by spectral leakage.

  \item First sidelobe width $\Omega_1$. Unlike $\Omega_0$, $\Omega_1$ is defined as the width of the first sidelobe peak. It is important to note that there is an identical first sidelobe at the negative frequency. Similar to the previous parameter $\widehat{W}_1$, this parameter measures the width of the spurious peak and characterizes the extent of energy into adjacent frequency bins. By utilizing $\widehat{W}_1$ and $\Omega_1$, we can estimate the amount of leaked energy contained in the first spurious peak.

  \item Normalized spectral decay scale $\Delta \Omega$. In most papers, the sidelobe roll-off ratio is often used to measure the rate of spectral leakage attenuation. We defined $\Delta \Omega$ as the frequency value corresponding to the maximum decibel of the sidelobe that is less than $-60dB$. It demonstrates the characteristic scale of the spectral leakage. Similar to the sidelobe roll-off ratio, a smaller $\Delta \Omega$ indicates better rejection of far-end interferences\cite{bergenDesignUltrasphericalWindow2004}. By utilizing $\Delta \Omega$, we can conveniently know the frequency where the spectral leakage intensity decays to 1/1000 of the original.

  \item Half width $\Delta$. In the time domain, this parameter measures the width of the window function at half of its maximum energy. It provides information about the temporal resolution and the extent to which the signal has not been modified.
\end{itemize}

By comparing the performance of different window functions using these parameters, we can assess their suitability for various applications and select the one that best meets the specific requirements of the problem at hand.

\begin{figure}[htb]
  \centering
  \includegraphics[height=0.30\textwidth]{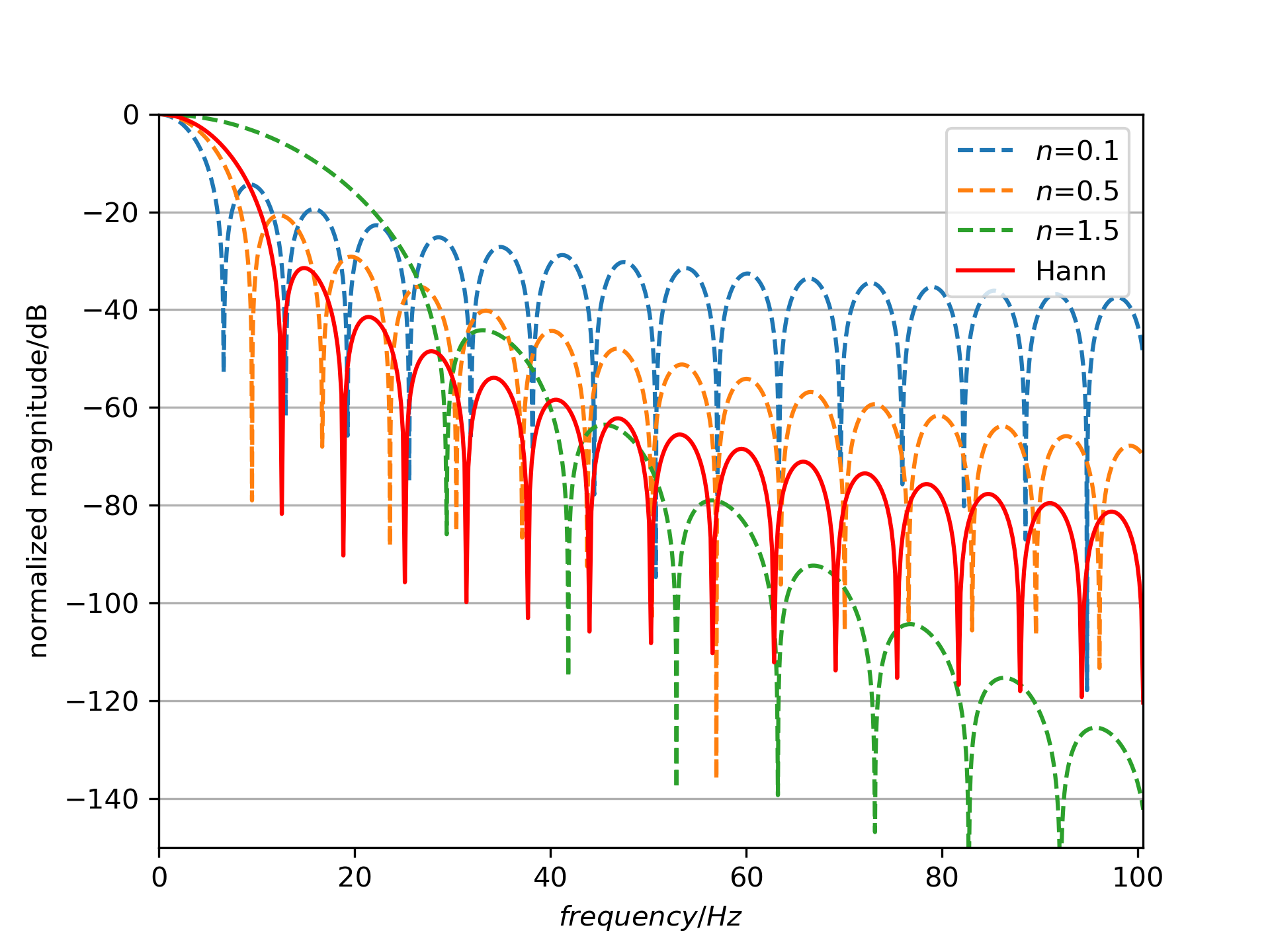} 
  \caption{Frequency spectrum with the polynomial $B(t)$ for different values of $n$.}
  \label{fig:FT_different_n}
\end{figure}

First, we investigate three cases for the symmetric polynomial $B(t)$ with different values of $n$: 0.1, 0.5, and 1.5 and examine their distributions in the Fourier space. As shown in Fig. \ref{fig:FT_different_n}, it is apparent that with an increase in the polynomial exponent $n$, the spectral decays more rapidly in the frequency domain. Simultaneously, there is an expansion in the width of the main lobe. This trade-off is a common characteristic in signal processing, where achieving both high frequency resolution and improved suppression of spectral leakage is challenging.

Furthermore, there exists another trade-off between time and frequency resolution. Window functions with higher values of $n$ exhibit greater concentration in the time domain, resulting in smaller halfwidths. However, they also tend to spread out more in the frequency domain, leading to larger main lobe widths. This trade-off reflects the uncertainty principle in signal processing. The faster decay of spectral with increasing frequency for higher values of $n$ indicates that window functions with higher values of $n$ offer better suppression of spectral leakage.

\begin{figure}[htb]
  \centering
  \includegraphics[height=0.30\textwidth]{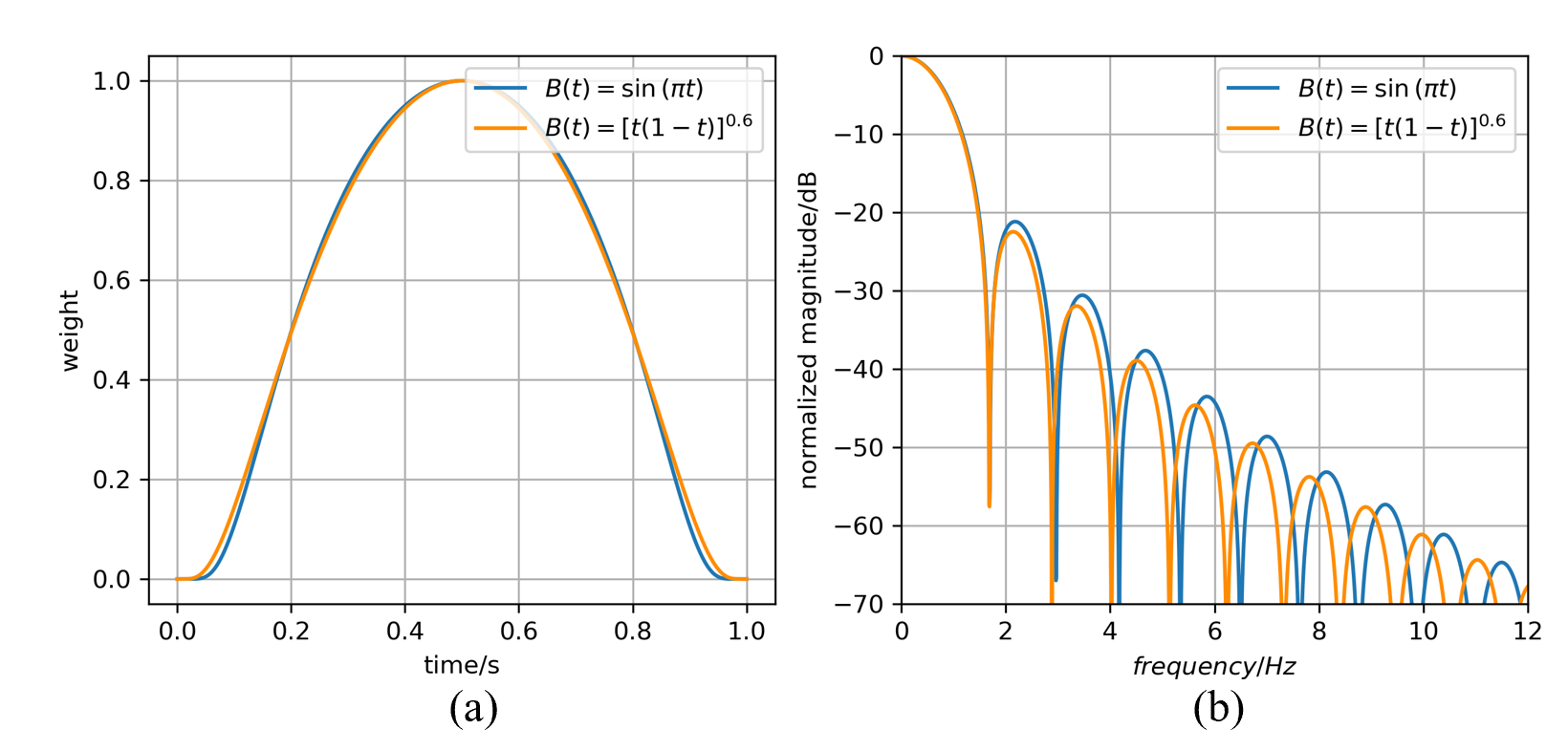}
  \caption{(a) Comparison of the window functions in the time domain. (b) Comparison of the window functions in the frequency domain.}
  \label{fig:similar_comparison}
\end{figure}

We notice that when $n=0.6$, the window function defined as $B(t) = t^n(1-t)^n$ exhibits a striking similarity to $\sin{(\pi t)}$, with only minor differences in their decay rates and width of sidelobes. The comparison of their distributions in the time and frequency domains is shown in Fig. \ref{fig:similar_comparison}. We examine their representations in the frequency domain and find remarkable similarities in terms of main lobe width, main lobe energy, and even the height of side lobes. This finding suggests the possibility of approximating more complex functions using polynomials. Furthermore, it indicates that by adjusting the exponent of the polynomial or combining different polynomials with varying exponents, we can easily construct more sophisticated and superior-performing window functions.

For many existing window functions, they exhibit a tendency to approach 0 from a positive value as they approach 0 or 1. This characteristic aligns with the desired behavior we seek for the function $B(t)$. Therefore, we can leverage this class of window functions to construct entirely new ones. As an illustration, we can begin with the sine window $\sin(\pi t)$ and enhance its continuity by constructing a window function (not normalized) as $\exp(-1/\sin(\pi t))$.

We can also introduce a coefficient to $B(t)$, such as $B(t) = 2\sin(\pi t)$ or $\sin(\pi t)/2$. Although such operations may seem trivial for general window functions, since $2\sin(\pi t)$ and $\sin(\pi t)$ have the same form when normalized, in our constructed class of functions, all coefficients appear as exponents in the window function. For instance, $2\sin(\pi t)$ corresponds to the window function $\left[\exp(-1/\sin(\pi t))\right]^{1/2}$. This introduces greater flexibility in constructing window functions by allowing for the inclusion of coefficients.

\begin{figure}[htb]
  \centering
  \includegraphics[height=0.30\textwidth]{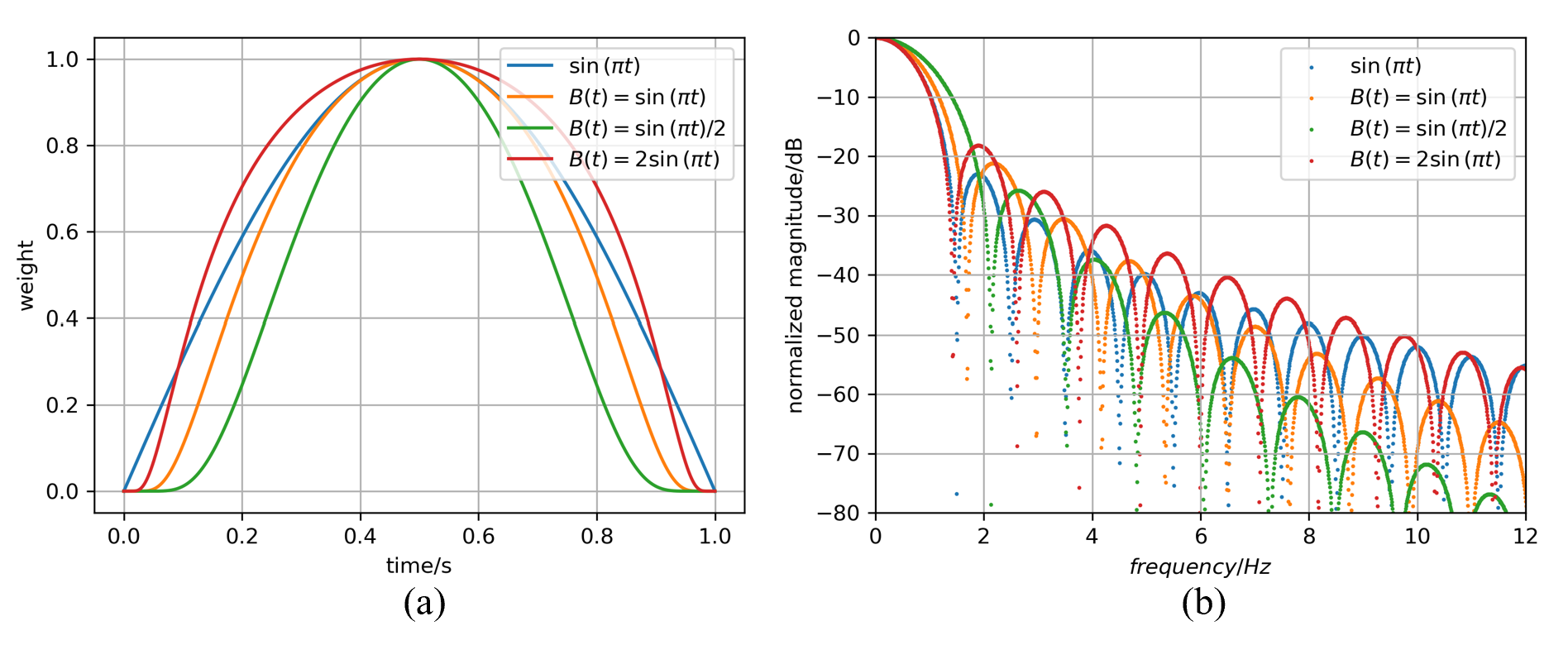} 
  \caption{(a) Comparison of the window functions in the time domain. (b) Comparison of the window functions in the frequency domain.}
  \label{fig:Construction}
\end{figure}

We analysis the differences between the sine window and three categories of constructed windows: $B(t) = \sin(\pi t)$, $2\sin(\pi t)$, and $\sin(\pi t)/2$. From Fig. \ref{fig:Construction}, we observe that our constructions not only enhance the smoothness around 0 and 1 compared to the original window function but also result in significant changes in the shape of the window function with different coefficients. In the frequency domain, we find that for our constructed window functions, as the coefficient increases, the main lobe width becomes narrower, but the decay scale and the height of the first sidelobe increases. However, irrespective of the coefficient used, these constructed windows exhibit faster decay compared to the original window function at large values of $\omega$. This indicates that our constructed class of window functions is more effective in controlling the scale of spectral leakage and managing the width of the main lobe.

We will calculate the performance parameters for the mentioned window functions and compare the results with other commonly used window functions. Rectangular (also known as Dirichlet window), Triangular (also known as Bartlet or Fejer window), and Welch windows\cite{harrisUseWindowsHarmonic1978,rice1964approximation,blackman1958measurement,welch1967use} are generated using the polynomial function $|t-0.5|^n$ centered on 0.5, which corresponds to the case where $B(t)$is polynomial. Sine and Hann windows\cite{harrisUseWindowsHarmonic1978,welch1967use} are typical representatives of sine/cosine window functions. We no longer calculate Blackman windows here, but Hamming window\cite{harrisUseWindowsHarmonic1978,welch1967use} as a comparison since Hamming window \cite{avciNewWindowBased2008a,1163349} is not zero at the endpoints but has great performance. Moreover, with the probability density function as the kernel, we can easily construct a window function centered on 0.5 and defined on $(0,1)$, such as Gaussian, Cauchy-Lorentz, and Poisson windows\cite{harrisUseWindowsHarmonic1978,achieser2013theory,bary2014treatise}. However, due to the characteristics of probability density functions, such a window function will not converge to zero at its endpoints. Kaiser window\cite{kaiser1964family,princeton1966system} is also widely used as representatives of window functions constructed using complex functions (in this case, modified Bessel functions of the first class). Tukey Windows and Planck-taper windows\cite{mckechan2010tapering,tukey1967intoroduction} are included as representatives of piecewise defined window functions.

For the analysis of the main lobe width, first sidelobe width, and spectral decay scale, our focus will be on the positive frequency part, and the results will be presented as $frequency \equiv \omega/2\pi(Hz)$. Window functions with star(*) indicates the ones that do not converge to zero at endpoints. For other window functions with adjustable parameters, such as Gaussian and Kaiser, the results will be computed using typical parameters. As for the Tukey and Planck-taper windows, which are typically segmented window functions, we will provide results for three sets of adjustable parameters. When the $1-I_0$ parameter is 0.00, it means that the energy concentrated in the main lobe exceeds 99.99\%, and there is almost no energy leakage. The results are presented in Table \ref{table}. 

\begin{sidewaystable}[ht]
  \begin{tabular}{cc|cccccc} 
    \hline\hline
      windows       &$B(t)/W(t)$            & $\Omega_0/2\pi Hz$ 
                                                 & $1-I_0/\%$ 
                                                      & $-\widehat{W}_1/dB$ 
                                                           & $\Omega_1/2\pi Hz$ 
                                                                & $\Delta \Omega/2\pi Hz$ 
                                                                     & $\Delta/0.1s$  \\ 
    \hline\hline
    \multirow{7}{*}{$e^{1/B(0.5)}e^{-1/B(t)}$} 
                    &Welch                  &1.59&1.01&20.1&1.23&11.2&5.07 \\
                    &Sine                   &1.69&0.76&21.2&1.27&10.4&4.67 \\
                    &$\frac{1}{2}\sin(\pi t)$
                                            &2.13&0.22&25.8&1.39&7.80&3.50 \\
                    &$2\sin(\pi t)$         &1.42&1.74&18.2&1.20&14.1&5.98 \\
                    &Hann                   &2.27&0.40&23.5&1.62&10.1&3.39 \\
                    &Kaiser($\alpha=8/\pi$) &2.70&0.24&25.4&1.87&10.0&2.80 \\
                    &Tukey$(\alpha=0.5)$    &1.40&4.87&14.3&1.40&13.3&6.69 \\
    \hline
    \multirow{6}{*}{$e^{4^n}e^{-1/t^n(1-t)^n}$} 
                    & n=0.1                 &1.05&6.51&14.4&1.01&140.6&9.63 \\
                    & n=0.25                &1.19&3.11&16.5&1.04&37.9&7.64 \\
                    & n=0.5                 &1.52&0.89&20.7&1.14&12.7&5.23 \\ 
                    & n=1.0                 &2.61&0.06&30.5&1.48&7.29&2.83 \\ 
                    & n=1.5                 &4.68&0.00&44.2&1.98&7.24&1.67 \\ 
                    & n=2.0                 &8.71&0.00&65.5&2.65&9.38&1.03 \\ 
    \hline\hline  
      Rectangular   &    1                  &1.00&9.71&13.3&1.00&317.5&10.0 \\
      Triangular    &   $1-|t-\frac{1}{2}|$ &2.00&0.29&26.5&2.00&21.0&2.93 \\
      Welch*
                    &   $4t(1-t)$           &1.43&0.79&21.3&1.03&18.0&5.41 \\
    \hline
      Sine          &   $\sin({\pi t})$       &1.50&0.51&23.0&1.00&16.0&5.00 \\
      Hann          &$0.5-0.5\cos(2\pi t)$  &2.00&0.05&31.5&1.00&7.46&3.64 \\ 
      Hamming       &$0.54-0.46\cos(2\pi t)$&2.00&0.04&44.1&0.60&47.5&3.82 \\ 
    \hline
    Gaussian*($\sigma=0.5$)
                    &$\exp\left(-\frac{1}{2}\left(\frac{t-0.5}{\sigma}\right)^2\right)$
                                            &1.11&4.48&16.5&0.95&225.5&8.32 \\
    Cauchy-Lorentz*($\gamma=0.5$)
                    &$\frac{\gamma^2}{(x-0.5)^2+\gamma^2}$
                                            &1.18&2.94&19.0&0.86&202.5&6.43 \\
    Poisson*($\tau=0.5$)
                    &$e^{-|t-0.5|/\tau}$    &1.30&2.19&25.5&0.61&185.5&3.47 \\
    \hline
    Kaiser($\alpha=8/\pi$)
                    &$\frac{I_0\left(\pi\alpha\sqrt{1-(2t-1)^2}\right)}{I_0(\pi\alpha)}$     
                                            &2.74&0.00&58.7&0.51&3.54&3.01  \\
    \hline
    Tukey($\alpha=0.3,0.5,0.7$)
                    & $\left\{\begin{aligned}
                      &\frac{1}{2}\left[1-\cos \left( 2\pi \frac{t}{\alpha}\right) \right],0<t<\frac{\alpha}{2}\\
                      &1,\frac{\alpha}{2}\leq t \leq 1-\frac{\alpha}{2}\\
                      &\frac{1}{2}\left[1-\cos \left( 2\pi \frac{1-t}{\alpha}\right) \right],1-\frac{\alpha}{2}<t<1
                    \end{aligned}\right.$ 
                                            &$\begin{aligned}
                                              &1.18\\
                                              &1.34\\
                                              &1.54
                                            \end{aligned}$
                                                 &$\begin{aligned}
                                                  &6.16\\
                                                  &3.75\\
                                                  &1.62
                                                \end{aligned}$
                                                      &$\begin{aligned}
                                                        &13.8\\
                                                        &15.1\\
                                                        &18.2
                                                      \end{aligned}$
                                                           &$\begin{aligned}
                                                            &1.17\\
                                                            &1.33\\
                                                            &1.54
                                                          \end{aligned}$
                                                                &$\begin{aligned}
                                                                  &9.67\\
                                                                  &9.63\\
                                                                  &4.45
                                                                \end{aligned}$
                                                                     &$\begin{aligned}
                                                                      &8.09\\
                                                                      &6.82\\
                                                                      &5.55
                                                                    \end{aligned}$
                                                                           \\
    Planck-taper($\varepsilon=0.15,0.25,0.35$)
                    & $\left\{\begin{aligned}
                      &\left[1+\exp \left( \frac{\varepsilon}{t}+\frac{\varepsilon}{t-\varepsilon}\right) \right]^{-1},0<t<\varepsilon\\
                      &1,\varepsilon\leq t \leq 1-\varepsilon\\
                      &\left[1+\exp\left( \frac{\varepsilon}{1-t}+\frac{\varepsilon}{1-t-\varepsilon}\right) \right]^{-1},1-\varepsilon<t<1
                    \end{aligned}\right.$ 
                                            &$\begin{aligned}
                                              &1.18\\
                                              &1.34\\
                                              &1.54
                                            \end{aligned}$
                                                &$\begin{aligned}
                                                  &6.96\\
                                                  &4.92\\
                                                  &2.86
                                                \end{aligned}$
                                                      &$\begin{aligned}
                                                        &13.6\\
                                                        &14.3\\
                                                        &16.0
                                                      \end{aligned}$
                                                          &$\begin{aligned}
                                                            &1.17\\
                                                            &1.33\\
                                                            &1.54
                                                          \end{aligned}$
                                                                &$\begin{aligned}
                                                                  &13.0\\
                                                                  &7.90\\
                                                                  &9.62
                                                                \end{aligned}$
                                                                    &$\begin{aligned}
                                                                      &8.18\\
                                                                      &6.97\\
                                                                      &5.76
                                                                    \end{aligned}$
                                                                          \\
    \hline\hline
  \end{tabular}
  \caption{Comparison of characteristic parameters of different window functions. $\Omega_0,1-I_0,\widehat{W}_1,\Omega_1,\Delta\Omega,\Delta$ indicate half main lobe width, energy leakage intensity,first sidelobe height, first sidelobe width, normalized spectral decay scale and half width.}
  \label{table}
  \end{sidewaystable}

Based on the results presented in Table \ref{table}, several observations can be made regarding the behavior of the tested window functions in the frequency domain. Except for the Kaiser window, most of the tested window functions exhibit an expansion in the main lobe width $2\Omega_0$ when reconstructed using the exponential construction method. Furthermore, all the tested window functions demonstrate an increase in the energy leakage intensity $1-I_0$, as well as an increase in the height $\widehat{W}_1$ and width $\Omega_1$ of the first sidelobe after reconstruction. When considering the characteristic scale of spectral leakage $\Delta \Omega$, which corresponds to the frequency at which the height of the sidelobe drops below -60 dB, we observe a decrease for the Welch and Sine windows, while Hann, Kaiser and Tukey windows display an increase in this measure. In the time domain, all the reconstructed window functions exhibit a reduction in their characteristic width $\Omega$ following modulation.

Upon examining the performance comparison between the original window functions and their reconstructed counterparts, it seems that the proposed method of constructing new window functions can not yield satisfactory results. However, let's shift our focus to three specific reconstructed window functions with $B(t)=\sin(\pi t)$, $\sin(\pi t)/2$, and $2\sin(\pi t)$. As mentioned earlier, adding an arbitrary constant coefficient to the window function itself does not have a meaningful effect since all coefficient-related information is lost after normalization. Nevertheless, in our proposed reconstruction scheme, the arbitrary constant coefficient of $B(t)$ appears as an exponent in the overall window function, thereby preserving the coefficient information.

When we increase the coefficient value, the main lobe width tends to decrease, which is advantageous for improving frequency resolution. However, this improvement comes at the cost of increased energy leakage. It is important to note that the increased energy leakage primarily results from an increase in the decay scale and the height of sidelobes, rather than the width of sidelobes. In the time domain, the characteristic width of the window function tends to increase, which can be beneficial for preserving more information in the data. These conclusions can be supported by comparing the results of $B(t)=t^n(1-t)^n$ with $n=1$ and the Welch function.

Exactly, in the special case of $B(t)=t^n(1-t)^n$, decreasing the exponent of the polynomial has a similar effect to increasing the coefficient of $B(t)$. In both scenarios, we observe similar trends in the behavior of the window function in both the time and frequency domains.

For problems with high SNR that require high frequency accuracy, it is advantageous to choose window functions with narrower main lobes. In such cases, we can conveniently increase the coefficient of the general $B(t)$ or decrease the exponent of the polynomial $B(t)$ to improve frequency resolution.

Conversely, when dealing with narrowband signals accompanied by strong interfering noise, where the requirement for main lobe width is reduced, we need window functions with higher decay rates and lower sidelobes to suppress interfering noise and enhance the SNR. In such cases, decreasing the coefficient of the general $B(t)$ or increasing the exponent of the polynomial $B(t)$ proves useful.

In the case of signals with extremely low SNR, such as gravitational wave signals in astronomy, it becomes crucial to select window functions that effectively preserve signal characteristics in the time domain. In these cases Tukey and Planck-taper windows are commonly used. Complementing these choices with Welch averaging for power spectrum computation can lead to favorable outcomes. In our proposed construction scheme, by carefully selecting an appropriate $B(t)$ function and adjusting the corresponding parameters, we can find suitable solutions tailored to address the challenges posed by these scenarios.

\section{Conclusion}

In this paper, we propose a new window function constructed using the exponential function. We conduct a comparative analysis of the performance of this class of window functions and other popular window functions in both the frequency domain and time domain by employing a high-resolution DFT method. Our findings demonstrate that the window functions developed in this study offer exceptional flexibility. By selecting appropriate kernel functions and adjusting coefficients, or utilizing polynomials with two free parameters (power and coefficient) as kernel functions, we can easily control the performance of the window functions. 

We believe that the introduction of this class of window functions in our paper presents a novel approach to constructing window functions with excellent performance. The various variations and remarkable flexibility of these window functions offer researchers in different fields a new control scheme for signal analysis. We also encourage further research on this window function to explore its application in various scenarios and fully unlock its potential.

\bibliographystyle{ieeetr}
\bibliography{reference.bib}

\end{document}